\documentclass{article}

\usepackage[latin2]{inputenc}
\usepackage{amsthm} 
\usepackage{amsmath} 
\usepackage{amssymb}
\usepackage[scanall]{psfrag}
\usepackage{graphics}
\usepackage{graphicx}

\theoremstyle{plain}

\theoremstyle{definition}

\newcommand{\startlist}{
\setlength{\itemsep}{0mm}
\setlength{\topsep}{2mm}
\setlength{\leftmargin}{4mm}
\setlength{\rightmargin}{0mm}
}
\newcommand{\listlabel}{$\bullet$}

\newenvironment{shortitemize}{
\begin{list}{\listlabel}{\startlist}
}{
\end{list}
}

\title{Large scale link based latent Dirichlet allocation for web
  document classification\thanks{Data Mining and Web search Research Group,
    Informatics Laboratory, Computer and Automation Research Institute
    of the Hungarian Academy of Sciences. e-mail: \{ibiro,
    jacint\}@ilab.sztaki.hu. This work was supported by the EU FP7
    project JUMAS -- Judicial Management by Digital Libraries
    Semantics and by grants OTKA NK 72845 and NKFP-07-A2
    \emph{TEXTREND}.}}  \author{ Istv\'an B\'{\i}r\'o \qquad J\'acint
  Szab\'o}

\begin{document}

\maketitle

\begin{abstract}
  In this paper we demonstrate the applicability of latent Dirichlet
  allocation (LDA) for classifying large Web document collections.
  One of our main results is a novel influence model that gives a
  fully generative model of the document content taking linkage into
  account.  In our setup, topics propagate along links in such a way
  that linked documents directly influence the words in the linking
  document.  As another main contribution we develop LDA specific
  boosting of Gibbs samplers resulting in a significant speedup in our
  experiments.  The inferred LDA model can be applied for
  classification as dimensionality reduction similarly to latent
  semantic indexing.  In addition, the model yields link weights that
  can be applied in algorithms to process the Web graph; as an example
  we deploy LDA link weights in stacked graphical learning.  By using
  Weka's BayesNet classifier, in terms of the AUC of classification,
  we achieve 4\% improvement over plain LDA with BayesNet and 18\%
  over tf.idf with SVM.  Our Gibbs sampling strategies yield about
  5-10 times speedup with less than 1\% decrease in accuracy in terms
  of likelihood and AUC of classification.
\end{abstract}

%\category{H.3}{Information Systems}{Information Storage and Retrieval}
%\category{I.2.7}{Computing Methodologies}{Artificial Intelligence}[Natural Language Processing]

%\terms{text analysis, feature selection, document classification, information retrieval}


\section{Introduction}
In this paper we demonstrate the applicability of latent Dirichlet
allocation \cite{blei2003lda}, a computationally challenging but very
powerful generative model for large scale Web document classification
relying on hyperlinkage in addition to text content.  Web content
classification is a research area that abounds with opportunities for
practical solutions.  The performance of most traditional machine
learning methods is limited by their disregard for the interconnection
structure between web data instances (nodes).  At the same time,
relational machine learning methods often do not scale to web-sized
data sets and, prior to our result, LDA models in general and in
particular those that leverage on the link structure were thought to
require an unfeasibly large amount of resources on the Web scale.

We apply one of the most successful generative topic models, latent
Dirichlet allocation (LDA) developed by Blei, Ng and Jordan
\cite{blei2003lda} for Web site classification.  Generative topic
models \cite{deerwester90indexing,hofmann2001ulp,blei2003lda} have a
wide range of applications
\cite{feifei2005bhm,mccallum2005art,bhattacharya2006ldm,wei2006lbd,xing2007eld}
in the fields of language processing, text mining and information
retrieval, including categorization, keyword extraction, similarity
search and statistical language modeling.  An LDA model consists of
latent topics described by distributions over vocabulary terms, and
every term occurrence arises based on the topic distribution
corresponding to the document in question.  As a starting point of our
results, we may use latent topics for dimensionality reduction prior
to classification as already suggested but since then less explored in
\cite{blei2003lda}.

Recently several models extend LDA to exploit links between web
documents or scientific papers
\cite{cohn2001mlp,erosheva2004mmm,dietz2007upc,nallapati:jlt}. In
these models the term and topic distributions may be modified along
the links.  All these models have the drawback that every document is
thought of either citing or cited, in other words, the citation graph
is bipartite, and influence flows only from cited documents to citing
ones.

In this paper we develop the \textbf{linked LDA} model, in which each
document can cite to and be cited by others and thus be influenced and
influence other documents.  Linked LDA is very similar to the copycat
model of Dietz, Bickel and Scheffer \cite{dietz2007upc} with the main
difference that in our case the citation graph is not restricted to be
bipartite. This fact and its consequences are the main advantage of
our model, namely, that the citation graph is homogeneous, and so one
does not have to take two copies, citing and cited, of every document.
In addition we give a flexible model of all possible effects,
including cross-topic relations and link selection.  As an example, we
may model the fact that topics Business and Computer are closer to one
another than to Health as well as the distinction between topically
related and unrelated links such as links to software to view the
content over a Health site.  The model may also distinguish between
sites with strong, weak or even no influence from its neighbors.  The
linked LDA model is described in full detail in Section~\ref{ss:llda}.

We demonstrate the applicability of linked LDA for text
categorization, an application although explicitly mentioned in
\cite{blei2003lda} but, in our best knowledge, justified prior to our
work only in special applications \cite{biro08lda}.  The inferred
topic distributions of documents are used as features to classify the
documents into categories. In the linked LDA model a weight is
inferred for every link.  In order to validate the applicability of
these edge weights, we show that their usage improves the performance
of stacked graphical classification, a meta-learning scheme introduced
in \cite{kou07stacked}.

The crux in the scalability of LDA for large corpora lies in the
understanding of the collapsed Gibbs sampler for inference.  In the
first application of the Gibbs sampler to LDA \cite{griffiths2004fst}
as well as in the fast collapsed Gibbs sampler \cite{porteous2008fcg}
the unit of sampling or, in other terms, a transition step of the
underlying Markov chain, is the redrawing of one sample for a single
term occurrence.  The storage space and update time of all these
counters prohibit sampling for very large corpora.  Since however the
order of sampling is neutral, we may group occurrences of the same
term in one document together.  Our main idea is then to re-sample
each of these term positions in one step and assign a joint storage
for them.  We introduce three strategies: for \textbf{aggregated}
sampling we store a sample for each position as before but update all
of them in one step for a word, for \textbf{limit} sampling we update
a topic distribution for each distinct word instead of drawing a
sample, while for \textbf{sparse} sampling we randomly skip some of
these words. All of these methods result in a significant speedup of
about 5-10 times, with less than 1\% decrease in accuracy in terms of
likelihood and AUC of classification.  The largest corpus where we
could successfully perform classification using these boostings
consisted of 100k documents (that is web sites with a total of 12M
pages), and altogether 1.8G term positions.

To assess the prediction power of the proposed features, we run
experiments on a host-level aggregation of the \texttt{.uk} domain,
which is publicly available through the Web Spam
Challenge~\cite{castillo08wsc}.  We perform topical classification
into one of 11 top-level categories of the Open Directory
(\texttt{http://dmoz.org}). Our techniques are evaluated along several
alternatives and, in terms of the AUC measure, yield the improvement
of 4\% over plain LDA and 18\% over tf.idf with SVM (here BayesNet is
used on linked LDA based features).

The rest of the paper is organized as follows. Section \ref{s:theory}
reviews the main concepts of LDA and then introduces our linked LDA
model. Section \ref{s:ex} describes the experimental setup and Section
\ref{s:res} the results.

\subsection{Related results}
The use of latent topics in information retrieval tasks starts with
latent semantic indexing (Deerwester, Dumais, Landauer, Furnas,
Harshman \cite{deerwester90indexing}), a method that represents
documents in a low rank approximation of the term space.
Probabilistic latent semantic analysis (PLSA, Hofmann
\cite{hofmann2001ulp}) extends this idea by defining a generative
model over the latent topics that yields a term distribution for each
document.  As the starting point of our results, latent Dirichlet
allocation (LDA, Blei, Ng, Jordan \cite{blei2003lda}) introduces
additional metaparameters and sampling from Dirichlet distributions to
yield a model with astonishing performance in various tasks.
% Lehetne kicsit szakszerubb, hogy mit is csinal az LDA a PLSA-hoz kepest.

We compare our linked LDA model to preexisting extensions of PLSA and
LDA that jointly model text and link as well as the influence of
topics along links. The first such model is PHITS defined by Cohn and
Hoffman \cite{cohn2001mlp}.  The mixed membership model of Erosheva,
Fienberg and Lafferty \cite{erosheva2004mmm} can be thought of as an
LDA based version of PHITS.  Common to these models is the idea to
infer similar topics to documents that are jointly similar in their
bag of words and link adjacency vectors.  Later, several similar link
based LDA models were introduced, including the copycat model, the
citation influence model by Dietz, Bickel and Scheffer
\cite{dietz2007upc} and the link-PLSA-LDA and pairwise-link-LDA models
by Nallapati, Ahmed, Xing and Cohen \cite{nallapati:jlt}.  These
results extend LDA over a bipartition of the corpus into citing and
cited documents such that influence flows along links from cited to
citing documents. They are shown to outperform earlier methods
\cite{dietz2007upc,nallapati:jlt}. The copycat model is very similar
to linked LDA, with the only difference that in the former every
document $d$ is duplicated into a citing and a cited copy, the topics
in the citing copy are drawn from $d$'s topic distribution, while
those in the cited copy are drawn from a cited document's topic
distribution. In contrast, in linked LDA, every topic either from $d'$
or a cited document's topic distribution. The citation influence model
is a finer version of the copycat model, in that there the citing
copy's topics are drawn either from $d'$ or a cited document's topic
distribution. The link-PLSA-LDA and pairwise-link-LDA models differ
from these in that they generate the links. We make comparisons to the
link-PLSA-LDA bipartite model in this paper.

While these four models generate topical relation for hyperlinked
documents, in a homogeneous corpus one has to duplicate each document
and infer two models for them.  This is in contrast to the linked LDA
model introduced in this paper whose main advantage is that it treats
citing and cited documents identically, and no duplication is needed.

As a completely different direction for link based LDA models, we
mention the results \cite{zhang2007pcd,zhang2007lbc,sinkkonen2008cml}
which give a generative model for the links of a network, with no
words at the nodes.

We also compare the performance of our results to general classifiers
aided by the Web hyperlinks.  Relational learning methods (presented,
for instance, in~\cite{getoor07statistical}) also consider existing
relationships between data instances.  The first relational learning
method designed for topical web classification was proposed by
Chakrabarti, Dom and Indyk \cite{Chakrabarti1998ehc} and improved by
Angelova and Weikum~\cite{angelova2006gbt}.  Several subsequent
results \cite[and the references therein]{qi06knowing} confirm that
classification performance can be significantly improved by taking
into account the labels assigned to neighboring nodes.  In our
baseline experiments we use the most accurate hypertext classifiers
\cite{castillo2006know} obtained by stacked graphical learning, a
meta-learning scheme introduced by Kou and Cohen \cite{kou07stacked}.
In stacked graphical learning, first a base learner is applied to the
training data to produce initial predictions.  Then the set of
features is expanded by adding the predictions of related instances
from the first step.  Finally, the base learner is re-applied to the
expanded feature set, resulting in a stacked model.  Performance of
stacked graphical learning is evaluated in Section~\ref{ss:baseline}
both with various graph based edge weights \cite{csalogany:ssl} and
with those inferred by our linked LDA model.

Another main contribution of this paper is three new efficient
inference methods.  Upon introducing LDA, Blei, Ng, Jordan
\cite{blei2003lda} proposed a variational algorithm for inference.
Later, several other methods were described for inference in LDA,
namely collapsed Gibbs sampling \cite{griffiths2004fst}, expectation
propagation, and collapsed variational inference \cite{teh2007cvb}.
Besides, \cite{newman2007dil} suggests methods how Gibbs sampling can
be applied in a paralleled environment.

Among the above collapsed Gibbs sampling methods, fas\-test
convergence is achieved by that of Griffiths and Steyvers
\cite{griffiths2004fst}.  To our best knowledge, prior to our result
there has been one type of attempt to speed up LDA inference in
general, and LDA-based Gibbs sampling in particular.  Porteous,
Newman, Ihler, Asuncion, Smyth, Welling \cite{porteous2008fcg} modify
Gibbs sampling such that it gives the same distribution by using
search data structure for sample updates, and call it fast Gibbs
sampler.  They show significant speedup for large topic numbers.

\section{LDA models and Gibbs sampling}\label{s:theory}

\subsection{Background}\label{ss:lda}

In order to prepare the necessary background and notation for our
linked LDA model in the next subsection, we shortly describe the Gibbs
sampling method for latent Dirichlet allocation \cite{blei2003lda}.
For a detailed elaboration we refer to Heinrich
\cite{heinrich2004pet}.  We have a vocabulary $V$ consisting of terms,
a set $T$ of $k$ topics and $m$ documents of arbitrary length. For
every topic $z\in T$ a distribution $\varphi_z$ on $V$ is sampled from
$\textrm{Dir}(\beta)$, where $\beta\in \mathbb R_+^V$ is a positive
smoothing parameter.  Similarly, for every document $d$ a distribution
$\vartheta_d$ on $T$ is sampled from $\textrm{Dir}(\alpha)$, where
$\alpha\in \mathbb R_+^T$ is a positive smoothing parameter.

\begin{figure}[ht]
  \centering
  \includegraphics[width=70mm]{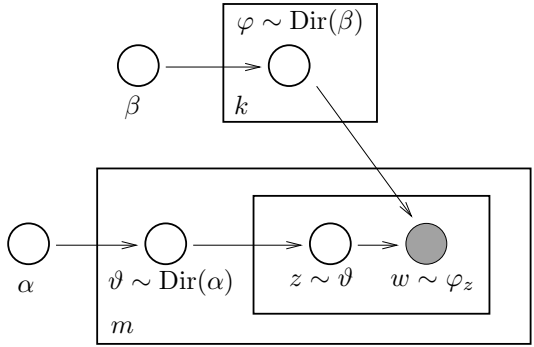}
  \caption{LDA as a Bayesian network}\label{figlda}
\end{figure}

The words of the documents are drawn as follows: for every word
position of document $d$ a topic $z$ is drawn from $\vartheta_d$, and
then a term is drawn from $\varphi_z$ and filled into that position.
The notation is summarized in the widely used Bayesian network
representation of LDA in Figure \ref{figlda}.

In this paper we use Gibbs sampling \cite{griffiths2004fst} for LDA
model inference.  Gibbs sampling is a Monte Carlo Markov chain
algorithm for sampling from a joint distribution $p(x),\,x\in \mathbb
R^n$, if all conditional distributions $p(x_i|x_{-i})$ are known
($x_{-i}= (x_1,\ldots, x_{i-1}, x_{i+1}, \ldots,x_n)$).  In LDA the
goal is to estimate the distribution $p(z|w)$ for $z\in T^P,\,w\in
V^P$ where $P$ denotes the set of word positions in the documents,
hence Gibbs sampling makes use of the values $p(z_i=z'|z_{-i},w)$ for
$i\in P$. In the initialization step a random topic assignment
$z_i,\,i\in P$ is chosen.

Gibbs sampling for LDA has an efficiently computable closed form as
deduced for example in \cite{heinrich2004pet}.  Before describing the
formula, we introduce the usual notation.  We let $d$ be a document
and $w_i$ its word at position $i$. We also let count $N_{dz}$ be the
number of words in $d$ with topic assignment $z$, $N_{zw}$ be the
number of words $w$ in the whole corpus with topic assignment $z$,
$N_d$ be the length of document $d$ and $N_z$ be the number of all
words with topic assignment $z$. A superscript $N^{-i}$ denotes that
position $i$ is excluded from the corpus when computing the
corresponding count.  Now the Gibbs sampling formula becomes
\begin{equation}\label{gibbs}
  p(z_i=z'|z_{-i},w) \propto
  \frac{N_{z' w_i}^{-i}+\beta(w_i)}{N_{z'}^{-i}+\sum_{w}\beta(w)} \cdot
  \frac{N_{d z'}^{-i}+\alpha(z')}{N_{d}^{-i}+\sum_{z}\alpha(z)}.
\end{equation}

After a sufficient number of iterations we stop with the current topic
assignment sample $z$. From $z$, the variables $\varphi$ and
$\vartheta$ are estimated as
\begin{equation}\label{phi}
\varphi_{z}(w)=\frac{N_{zw}+\beta(w)}{N_{z}+\sum_{w\in V}\beta(w)},
\end{equation}
and
\begin{equation}\label{theta}
\vartheta_{d}(z)=\frac{N_{dz}+\alpha(z)}{N_{d}+\sum_{z\in T}\alpha(z)}.
\end{equation}

The likelihood of an inferred LDA model on a set $P$ of word positions
in a collection of held-out documents is
\begin{equation}\label{ldalik}
  \prod_{i\in P} p(w_i)^{-1/|P|} \mbox{\quad where } \,\,\, p(w_i) =
  \sum_{z\in T}\varphi_{z}(w_i)\vartheta_{d}(z),
\end{equation}
where $d$ is document of position containing $i$. Here $\vartheta$ is
a MAP estimate of the topic-distribution of the document, and is
usually approximated by unseen inference.

\subsection{Linked LDA}\label{ss:llda}
\begin{figure}[ht]
  \centering
  \includegraphics[width=80mm]{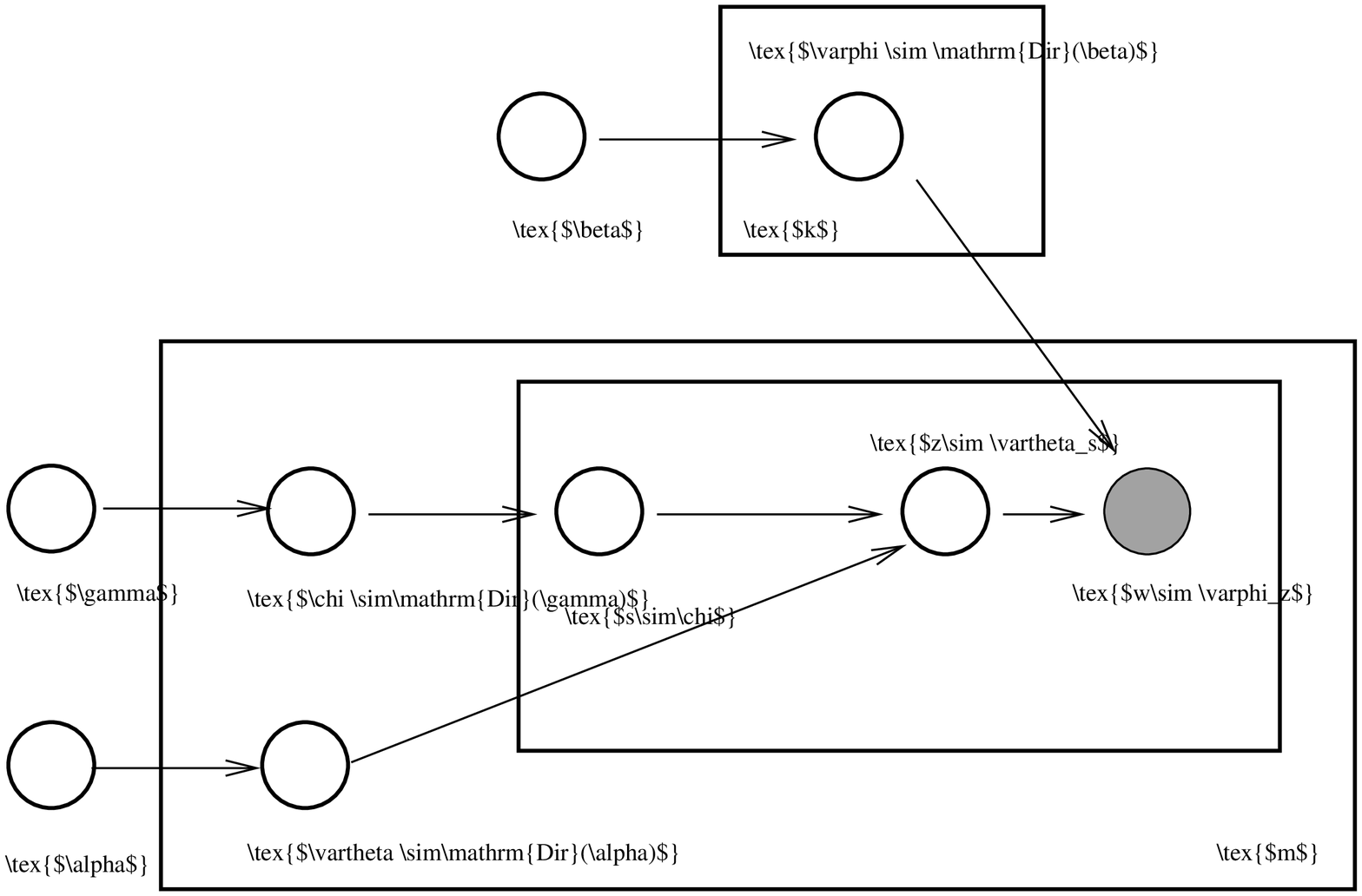}
  \caption{Linked LDA as a Bayesian network}\label{figlinkedlda}
\end{figure}

Next we extend latent Dirichlet allocation to model the effect of a
hyperlink between two documents on topic and term distributions.  The
key idea, summarized as a Bayes net in Figure \ref{figlinkedlda}, is
to modify the topic distribution of a position on the word plate based
on a link from the current document on the document plate.  For each
position we select either an outlink or the document itself to modify
the topic distribution of the original LDA model.

Formally we introduce linked LDA over the notations of the previous
subsection for vocabulary $V$, the $k$-element topic set $T$ and the
document set $D$.  Links are represented by a directed graph with
inlinks for cited and outlinks for citing documents.  Our model also
relies on the LDA distributions $\varphi_z$ and $\vartheta_d$.  We
introduce an additional distribution $\chi_d$ on the set $S_d = \{d$
and its outneighbors$\}$ for every document $d$, sampled from
$\textrm{Dir}(\gamma_d)$, where $\gamma_d$ is a positive smoothing
vector on $S_d$.

As also seen in the Bayes net of Figure~\ref{figlinkedlda}, the words
of the documents are drawn as follows.  For every word position $i$ of
document $d$, we
\begin{shortitemize}
\item draw an \textbf{influencing document} $r\in S_d$ from $\chi_d$,
\item draw a topic $z$ from $\vartheta_r$ (instead of $\vartheta_d$ as in LDA),
\item draw a term from $\varphi_z$ and fill into the position.
\end{shortitemize}

Note that for sake of a unified treatment, $d$ itself can be an
influencing document of itself. This is in contrast to the citation
influence model of \cite{dietz2007upc}, where for every word a
Bernoulli draw decides whether in the citing copy the influencing
document is $d$ itself or an outneighbor of it.

We describe the Gibbs sampling inference procedure for linked LDA.
Naturally, here $N_{dz}$ denotes the number of words with topic
assignment $z$ \emph{influenced by} document $d$, and similarly for
$N_{zw}, N_{d}$ and $N_{z}$. Note that this document $d$ is not
necessarily the one containing word $w$, it can be an outneighbor as
well. The goal is to estimate the distribution $p(r,z|w)$ for $r\in
D^P, z\in T^P,\,w\in V^P$ where $P$ denotes the set of word positions
in the documents. In Gibbs sampling one has to calculate
$p(z_i=z',r_i=r'|z_{-i},r_{-i},w)$ for $i \in P$. First, it can be
shown that
\begin{eqnarray}\label{gibbs2}
  p(r,z,w|\varphi,\vartheta,\chi) &=&\\
  &&\hspace{-2cm}\prod_{d\in D} \frac{\Delta(\mathbf{M}^r_d+\gamma_d)}{\Delta(\gamma_d)} \cdot
  \prod_{d\in D} \frac{\Delta(\mathbf{N}^z_d+\alpha)}{\Delta(\alpha)} \cdot
  \prod_{z\in T} \frac{\Delta(\mathbf{N}^w_z+\beta)}{\Delta(\beta)}.\notag
\end{eqnarray}
Here $\mathbf{M}^r_d$ is the count vector of influencing documents
appearing at the words of document $d$, $\mathbf{N}^z_d$ is the count
vector of observed topics influenced by $d$, and $\mathbf{N}^w_z$ is
the count vector of words with topic $z$.

By division, \eqref{gibbs2} gives an iteration in the Gibbs sampling as
follows:
\begin{eqnarray}
  p(z_i=z',r_i=r'|z_{-i},r_{-i},w) &=&\\\notag
  &&\hspace{-2cm}\frac{p(z,r,w)}{p(z_{-i},r_{-i},w_{-i})} \cdot 
     \frac{1}{p(w_i | z_{-i},r_{-i},w_{-i})},\notag
\end{eqnarray}
thus
\begin{eqnarray}\label{2}
  p(z_i=z',r_i=r'|z_{-i},r_{-i},w) &\propto&\\
 \notag \frac{p(z,r,w)}{p(z_{-i},r_{-i},w_{-i})} &=& \\\notag
  &&\hspace{-3.2cm}\frac{N^{-i}_{r'z'}+\alpha(z')}   {N^{-i}_{r'}+\sum_{z\in T}\alpha(z)} \cdot
  \frac{M^{-i}_{dr'}+\gamma_d(r')} {\sum_{r\in S_d}(M^{-i}_{dr}+\gamma_d(r))} \cdot\\\notag
  &&\hspace{0.1cm}\frac{N^{-i}_{z'w_i}+\beta(w_i)}    {N^{-i}_{z'}+\sum_{w\in V}\beta(w)}.\notag
\end{eqnarray}
Here $M_{dr}$ denotes the number of positions in document $d$
influenced by outneighbor $r\in S_d$.

%Ez a koordinatankenti Gibbs sampling (nem duplan hanem 2x szimplan samplingelunk):
% It follows that,
% \begin{eqnarray}\label{1z}
%   p(z_i=z'|z_{-i},r,w)  &\propto&\\
%   &&\hspace{-3cm}\frac{N^{-i}_{z'r_i}+\alpha(z')}   {\sum_{z\in T}(N^{-i}_{zr_i}+\alpha(z))-1} \cdot%\\\notag
%   \frac{N^{-i}_{z'w_i}+\beta(w_i)}    {\sum_{w\in V}(N^{-i}_{z'w}+\beta(w))-1},\notag
% \end{eqnarray}
% and
% \begin{eqnarray}\label{1r}
%   p(r_i=r'|z,r_{-i},w) &\propto&\\
%   &&\hspace{-3cm}\frac{N^{-i}_{r'z_i}+\alpha(z_i)}   {\sum_{z\in T}(N^{-i}_{r'z}+\alpha(z))-1} \cdot%\\\notag
%   \frac{N^{-i}_{dr'}+\gamma_d(r')} {\sum_{r\in S_d}(N^{-i}_{dr}+\gamma_d(r))-1} \notag
% \end{eqnarray}
% where $d_i$ is the containing document, $r_i$ is the influencing document,
% $z_i$ is the topic and $w_i$ is the word of position $p$.

% We compare two sampling methods.  We may sample on both $z$ and $r$
% at the same time as in Equation \ref{2}.  We refer to this method as
% \textbf{double} sampling.  Or we may sample on $z$ and $r$ one after
% another as in Equations \ref{1z} and \ref{1r} referred to as
% \textbf{simple} sampling.

While this ``two-coordinate'' sampling is against the general Gibbs
procedure that re-samples one coordinate at a time, it has the
required distribution $p(z|w)$ as its unique stationary
distribution. This follows from the fact that it is a random walk over
a finite state Markov chain which is irreducible and aperiodic as
$\alpha, \beta >0$. Thus it has a unique stationary distribution,
which is necessarily the distribution $p(z|w)$ by construction.

Similarly to LDA, after a sufficient number of iterations we stop with
the current topic assignment sample $z$, and estimate $\varphi$ as in
\eqref{phi}, $\vartheta$ as in \eqref{theta}, and $\chi$ by
\begin{equation}\label{chi}
  \chi_d(r)=\frac{M_{dr}+\gamma_{d}(r)}{\sum_{r\in S_d}(M_{dr}+\gamma_{d}(r))}.
\end{equation}

% Kihagytuk, mert bonyolult lett volna magyarazni. Istvan az
% osztalyozasokat csinalta ezzel, es azonos eredmenyeket adott
% \theta-val.
%% Note that
%% \begin{equation}\label{theta'}
%%  \vartheta'_{d}(z)=\frac{N'_{dz}+\alpha(z)}{N'_{d}+\sum_{z}\alpha(z)},
%% \end{equation}
%% where $N'_{dz}$ is the number of observed topics \emph{in the
%%   positions} of $d$, is an alternative empirical topic distribution of
%% document $d$.

For an unseen document $d$, distributions $\vartheta$ and $\chi$ can
be estimated exactly as in \eqref{theta} and \eqref{chi} once we have
a sample from its word topic assignment $z$ and word influencing
document assignment $r$. Similarly to LDA, unseen inference for linked
LDA is the same as doing linked LDA model inference for the whole
corpus (train and test corpora), in such a way that the $z$ and $r$
assignments for the training documents are kept throughout.

The likelihood of the inferred linked LDA model on a held-out corpus
is calculated analogously as for LDA in Equation \eqref{ldalik}:
\begin{equation}\label{linkedlik}
\prod_{i\in P} p(w_i)^{-1/|P|} \mbox{\quad where } p(w_i) =
\sum_{z\in T, r\in S_d}\varphi_{z}(w_i)\vartheta_{r}(z)\chi_{d}(r).
\end{equation}

\subsection{Fast Gibbs sampling heuristics}\label{ss:gibbs}
In this section we describe three strategies for faster inference for
both LDA and linked LDA.  The methods modify the original Gibbs
sampling procedure \cite{griffiths2004fst}. For simplicity, we
describe them for the plain LDA setting.  The speedup obtained by
these boostings is evaluated in Section~\ref{ss:speedup}.

All of our methods start by sorting the words of the original
documents so that sampling is performed subsequently for the
occurrences of the same word.  We introduce additional heuristics to
compute the new samples for all occurrences of the same word in a
document at once. 

In \textbf{aggregated} Gibbs sampling we calculate the conditional
topic distribution $F$ as in Equation \eqref{gibbs} for the first
occurrence $i$ of a word in a given document.  Next we draw a topic
from $F$ for \emph{every} position with the same word without
recalculating $F$ and update all counts corresponding to the same
word. In this way the number of calculations of the conditional topic
distributions is the number of different terms in the document instead
of the length of the document, moreover, the space requirement remains
unchanged. Thus the speedup is larger if there are more multiple
occurrences of the words. This mostly happens for large corpora.  This
performance can be further improved by maintaining the aggregated
topic count vector for terms with big frequency in the document,
instead of the storing the topic at each word.

\textbf{Limit} Gibbs sampling heavily relies on the bag of words model
assumption that the topic of a document remains unchanged by
multiplying all term frequencies by a constant.  In the limit hence we
may maintain the calculated conditional topic distribution $F$ for the
set of all occurrences of a word, without drawing a topic for every
single occurrence. Equation~\eqref{gibbs} can be adapted for this
setting by a straightforward redefinition of the $N$ counts.  

It is easy to check that if the topic distributions for all positions
are uniform then we get an instable fixed point of limit Gibbs
sampling, provided both $\alpha$ and $\beta$ are constant. Clearly,
with large probability, these fixed points can be avoided by selecting
biased initial topic distributions. We never encountered such instable
fixed points during our experiments.

Similarly to aggregated Gibbs sampling, depending on the size and term
frequency distribution of the documents, limit sampling may result in
compressed space usage.
% We use limit Gibbs sampler only in conjunction with the aggregated
% sampler, as it did not provide significant speedup used alone.

\textbf{Sparse} sampling with sparsity parameter $\ell$ is a lazy
version of limit Gibbs sampling where we ignore some of the less
frequent terms to achieve faster convergence on the more important
ones. On every document we sample doclength$/\ell$ times from a
multinomial distribution on the distinct terms with replacement, by
selecting a term by a probability proportional to its term frequency
tf$_w$ in the document.  Hence with $\ell=1$ we expect a performance
similar to limit Gibbs sampling, while large $\ell$ results in a
speedup of about a factor of $\ell$, with a trade-off of lower
accuracy and slower convergence. The idea of laziness can be naturally
built upon aggregated sampling alone, without limit sampling, and we
will indeed evaluate this sampling (called \textbf{aggregated sparse})
in Table \ref{tabbig} in Section \ref{ss:baseline}.

Aggregated Gibbs sampling has the required distribution $p(z|w)$ as
its unique stationary distribution. Indeed, it is a random walk over a
finite state Markov chain which is irreducible and aperiodic as
$\alpha, \beta >0$, implying that it has a unique stationary
distribution, which is necessarily the distribution $p(z|w)$ by
construction.

As for limit Gibbs sampling, we rely on the assumption, that in the
bag of words model, multiplying all term frequencies in the corpus by
a constant the semantic meaning of the documents change only
moderately. As limit Gibbs sampling arises as a limit of aggregated
Gibbs sampling by tending this constant to infinity, its stationary
distribution is very close to what the aggregated version samples,
that is, the required $p(z|w)$. This argument is justified by our
measurements in Section \ref{s:res}.

Laziness clearly keeps the above arguments valid, thus aggregated
sparse Gibbs sampling has stationary distribution $p(z|w)$, and sparse
sampling the same as limit.

\section{Experimental setup}\label{s:ex}

\subsection{The data set and classifiers}\label{ss:corpvocab}
In our experiments we use the 114k node host-level aggregation of the
WEBSPAM-UK2007 \texttt{.uk} domain crawl, which consists of 12.5M pages
and is publicly available through the Web Spam
Challenge~\cite{castillo08wsc}.  We perform topical classification
into one of the 14 top-level English language categories of the Open
Directory (\texttt{http://dmoz.org}) while excluding category ``World''
containing non-English language documents.  If a site contains a page
registered in DMOZ with some top category, then we label it with that
category. In case of a conflict we choose a random page of the site
registered in DMOZ and label its site with its top category. In this
way we could derive category label for 32k documents.
% Lehet, hogy nem csak words hanem meta keywords is, Pereszlenyi
% Attila adta a warcot ibironak, o tudja.

We perform the usual data cleansing steps prior to classification.
After stemming by
TreeTagger\footnote{\texttt{http://www.ims.uni-stuttgart.\-de/\-projekte/\-corplex/\-TreeTagger/}}
and removing stop words by the Onix
list\footnote{\texttt{http://www.lextek.\-com/\-manuals/\-onix/\-stopwords1.html}},
we aggregate the words appearing in all HTML pages of the sites to
form one document per site.  We discard rare terms and keep the 100k
most frequent to form our vocabulary.  We discard all hosts that
become empty, that is those consisting solely of probably only a few
rare terms.  To reduce unnecessary computational load we also discard
all unlabeled hosts with more than 100k remaining word occurrences.
Finally we weight directed links between hosts by their multiplicity.
For every site we keep only at most 50 outlinks with largest weight.

We call this the \textbf{big corpus}, as it contains 1.8G positions,
far more than the usual corpora on which LDA experiments are carried
out. Since the big corpus is infeasible for the baseline experiments,
we also form a \textbf{small corpus} consisting of the labeled hosts
only, to be able to compare our results with the baseline.  We use the
most frequent 20k terms as vocabulary and keep only the 10 largest
weight outlink for each host. Note that even the small corpus is large
enough to cause efficiency problems for the baseline classifiers.

We chose 11 out of the 14 categories to apply classification to them,
as the other 3 categories were very small in our corpus.  In our
experiments we perform two-class classification for all of these 11
big categories.  We use the machine learning toolkit Weka \cite{weka}
to apply SVM, C4.5 decision tree and the Bayes net implementation of
Weka, called BayesNet. As for graph stacking we used a home made Java
code integrated into Weka. We use 10-fold cross validation and measure
performance by the average AUC over the 11 classes. Every run
((linked) LDA model build and classification) is repeated 10 times to
get variance of the AUC classification performance.

\subsection{Baseline classifiers}\label{ss:class}
As the simplest baseline we use the tf.idf vectors with SVM for the
small corpus, as it took a prohibitively long time to run it on the
big corpus. Another baseline is to use the LDA delivered $\vartheta$
topic distributions as features with the classifier BayesNet.

As recently a large number of relational learning methods were
invented, a complete comparison is beyond the scope of this paper.
Instead we concentrate on the stacked graphical learning method
\cite{kou07stacked} that reaches best performance for classifying Web
spam over this corpus \cite{castillo2006know}.

The general stacked graphical procedure starts with one of the base
learners of Subsection~\ref{ss:corpvocab} that classifies each element
$v$ positive with weight $p (v)$.  Positive and negative instances in
the training set have $p (v)$ equal 0 and 1, respectively.  These
values are used in a classifier stacking step to form new features $f
(u)$ based on certain $p (v)$.  Stacking can recursively be applied,
hence if in one step we consider $p (v)$ for the neighbors of $u$,
then in a two-layer stacking we gather information from the distance
two neighborhood.

We use \textbf{cocitation} to measure node similarities.  The
cocitation coc$(u, v)$ is defined as the number of common inneighbors
of $u$ and $v$. This measure turned out most effective for Web spam
classification \cite{benczur06linkbased}.  We may use both the input
directed graph and its transpose by changing the direction of each
link.  We will refer to these variants as \textbf{directed} and
\textbf{reversed} versions.  Notice that reversed cocitation denotes
bibliographic coupling (nodes pointed to by both $u$ and $v$). Several
other options to measure node similarities and form the neighborhood
aggregate features are explored in
\cite{benczur06linkbased,csalogany:ssl}.

\subsection{LDA inference}\label{ss:tlda}
In our LDA inference the following parameter settings are used.  The
number of topics is chosen to be $k=30$.  The Dirichlet parameter
vector $\beta$ is constant $200/|V|$, and $\alpha$ is constant $50/k$.
For linked LDA we also consider directed links between the documents.
For a document $d$, the smoothing parameter vector $\gamma_d$ was
chosen in such a way that
\[\gamma_d(c) \propto w(d \to c)\,\,\,\mbox{ for all } c\in S_d, c\neq
d \mbox{ and}\]
\[\gamma_d(d) \propto 1+\sum_{c\in S_d, c\neq d} w(d \to c)\]
such that $\sum_{c\in S_d} \gamma_d(c) = |d| / p$, where $|d|$ is the
number of word positions in $d$ (the document length), and $w(d \to
c)$ denotes the multiplicity of the $d \to c$ link in the corpus. As a
quick parameter sweep, we tried three values $p=1,4,10$ with plain
Gibbs sampler and BayesNet classifier. The accuracy was 0.835, 0.852
and 0.854 resp., so we chose $p=10$ in the subsequent experiments.

We take the $\vartheta$ topic distribution vectors as features for the
documents and use the classifiers of Subsection~\ref{ss:corpvocab}.
For Gibbs sampling in LDA and linked LDA we apply the baseline as well
as the aggregated, limit and sparse heuristics.  Altogether this
results in eight different classes of experiments.

As another independent experiment we also measure the quality of the
inferred edge weight function $\chi$.  This weight can be used in the
stacked graphical classification procedure of Section~\ref{ss:class}
over the $\vartheta$ topic distribution feature and the tf.idf
baseline classifiers, for both the link graph and its reversed
version.

The same experiments are performed with the link-PLSA-LDA model
\cite{nallapati:jlt}, using the C-code provided to us by Ramesh
Nallapati.  We also compared the running time of our Gibbs sampling
strategies with the fast Gibbs sampling method of
\cite{porteous2008fcg}, using the C-code referred to
therein\footnote{\texttt{http://www.ics.uci.edu/\~{}iporteou/fastlda/}}.
For results, see Subsection \ref{ss:speedup}.

We developed an own C++-code for LDA and linked LDA containing plain
Gibbs sampling and the three Gibbs sampling boostings proposed in this
paper. This code is publicly
available\footnote{\texttt{http://www.ilab.sztaki.hu/\~{}ibiro/linkedLDA/}}
together with the used DMOZ labels of the \texttt{.uk} sites.
% As part of the design for high efficiency, our topic-word count table
% contains $N_{z w}+\beta_w$ instead of simply $N_{z w}$, and similarly,
% $\alpha$ and $\gamma$ are integrated into to the document-topic and
% document-influencing document count vectors. The use of this data
% representation resulted in XX\% speedup in Gibbs sampling.
% camera readybe beirni, addigra istvan is implementalja
The computations were run on Linux machines with 50GB RAM and
multicore 1.8GHz AMD Opteron processors with 1MB cache.

\section{Results}\label{s:res}
\subsection{Speedup with the Gibbs sampling strategies}\label{ss:speedup}
Applying aggregated, limit and sparse Gibbs samplings results in an
astonishing speedup, see Tables \ref{tspeedup1}-\ref{tspeedup2}. Experiments were
carried out on the small corpus, and the models were run with $k=30$
topics. 
% As linked LDA with limit Gibbs sampling did not fit into 100Gb
% memory with 30 topics, we run it with $k=10$ topics instead, and
% tripled the resulting running time of one iteration.  This workaround
% is fully justified by the fact that the running time is linear in the
% topic-number for the linked LDA model with limit Gibbs sampling (and
% also in the total number of different terms in the documents).
%megsem kellett igy

\begin{table}[ht]
\begin{center}
  \begin{tabular}{|l|c|c|c|}
    \hline
    Gibbs sampler&LDA&linked LDA\\
    \hline
    plain&1000&1303\\
    aggregated&193&1006\\
    limit&190&970\\
    sparse ($\ell=2$)&135&402\\
    sparse ($\ell=5$)&105&241\\
    sparse ($\ell=10$)&91&171\\
    sparse ($\ell=20$)&84&129\\
    sparse ($\ell=50$)&80&107\\
    \hline
\end{tabular}
\caption{\label{tspeedup1} Average CPU times for one iteration of the
  Gibbs sampler (in secs)}
\end{center}
\end{table}

\begin{table}[ht]
\begin{center}
  \begin{tabular}{|l|c|c|c|}
    \hline
    model / sampler&time\\
    \hline
    LDA / fast Gibbs \cite{porteous2008fcg}&949\\
    link-PLSA-LDA / var. inf. \cite{nallapati:jlt}&19,826\\
    \hline
\end{tabular}
\caption{\label{tspeedup2} Average CPU times for one iteration for
  baseline models and samplers (in secs)}
\end{center}
\end{table}

% A link-PLSA-LDA-nal a 74h ugy jott ki, hogy egy 35M poz szamu, 5.5M
% kuapcszamu corpuson k=30-cal 202min alatt futott le Istvannak 5
% iter. Ez 2424sec per iteracio, es a mi corpusunk 3.86G poz-t tart.
% Tehat ha a kupacszamban linearis a var inf, ami egyaltalan nem
% biztos, ez kb 74h lenne a mi corpusunkon.

Observe that the speedup of aggregated Gibbs sampling is striking, and
this does not at all come at the expense of lower accuracy. Indeed,
results of Subsections \ref{ss:lik} and \ref{ss:baseline} show less
than 1\% decrease in terms of likelihood and AUC value after 50
iterations when using aggregated Gibbs sampling. The advantage of
limit sampling over aggregated sampling (no need to draw from the
distribution) turns out to be negligible.  For sparse Gibbs sampling
the speedup is a straightforward consequence of the fact that we skip
many terms during sampling, and thus the running time is approximately
a linear function of $1/\ell$ where $\ell$ is the sparsity parameter.
The constant term in this linear function is apparently approximately
75sec for LDA and linked LDA, certainly, this is the time needed to
iterate through the 16G memory. As for the choice $\ell=10$, note that
the running time of one iteration for LDA is only 9\% of the one with
plain Gibbs sampling, and still, the accuracy measured in AUC is only
2\% worse as seen in Table \ref{tabsmall}.
%Nem biztos hogy ez a 75seces a magyarazat.

We point out that though we used the same C-code than
\cite{porteous2008fcg}, our measurement on fast Gibbs sampling
demonstrates a somewhat poorer performance than the results presented
in \cite{porteous2008fcg}, even if one takes into account that fast
Gibbs sampling is proven to have better performance with a large
number of topics ($k \approx 1000$) (with $k=100$ topics we
experienced 2100sec on average). On the other hand, fast Gibbs sampler
gets faster and faster for the consecutive iterations: in our
experiments with $k=30$ topics we measured 2656sec for the first
iteration (much worse than for plain Gibbs) and 757sec for the
$50^{\mbox{\small{th}}}$.  Thus the calculated average would be better
with more iterations -- to which, however, there is no real need by
the observations in Subsection \ref{ss:lik}.
% The same running times with $k=100$ topics is 8385 / 1632 secs.

As the number of iterations with variational inference is usually
chosen to be around 50, the same as for Gibbs sampling, we feel the
above running times for Gibbs sampling and link-PLSA-LDA with its
variational inference are comparable.

\subsection{Likelihood and convergence of LDA inference}\label{ss:lik}
\begin{figure}[ht]
  \centering
  \hspace{-.1cm}\includegraphics[width=60mm,angle=-90]{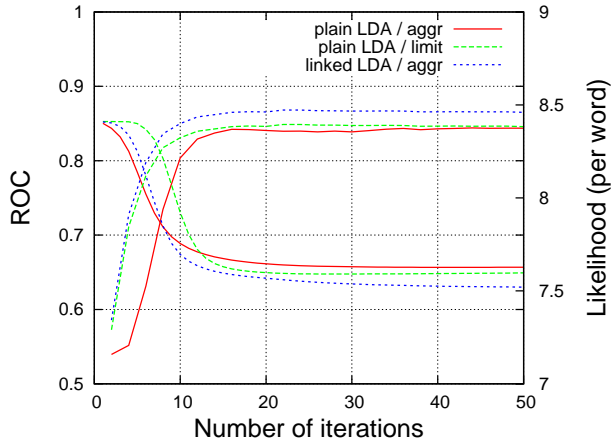}
  \caption{Correlation of the likelihood (the lower the better) and
    the AUC for BayesNet (the higher the better) for three choice of
    model / sampler combinations}\label{lik1}
\end{figure}

\begin{figure}[ht]
  \centering
  \hspace{-.1cm}\includegraphics[width=60mm,angle=-90]{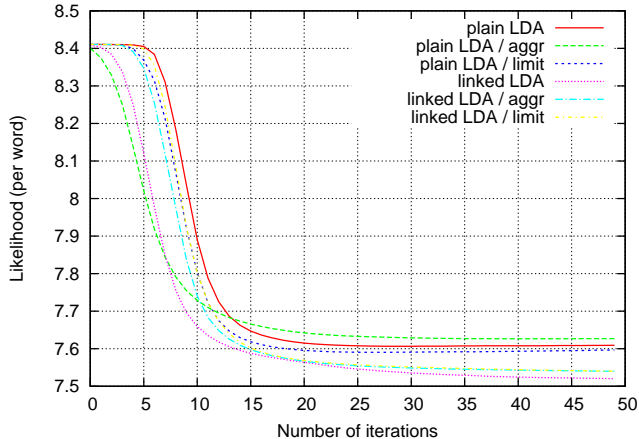}
  \caption{Convergence of the likelihood for various models and
    samplers}\label{lik2}
\end{figure}

\begin{figure}[ht]
  \centering
  \hspace{-.1cm}\includegraphics[width=60mm,angle=-90]{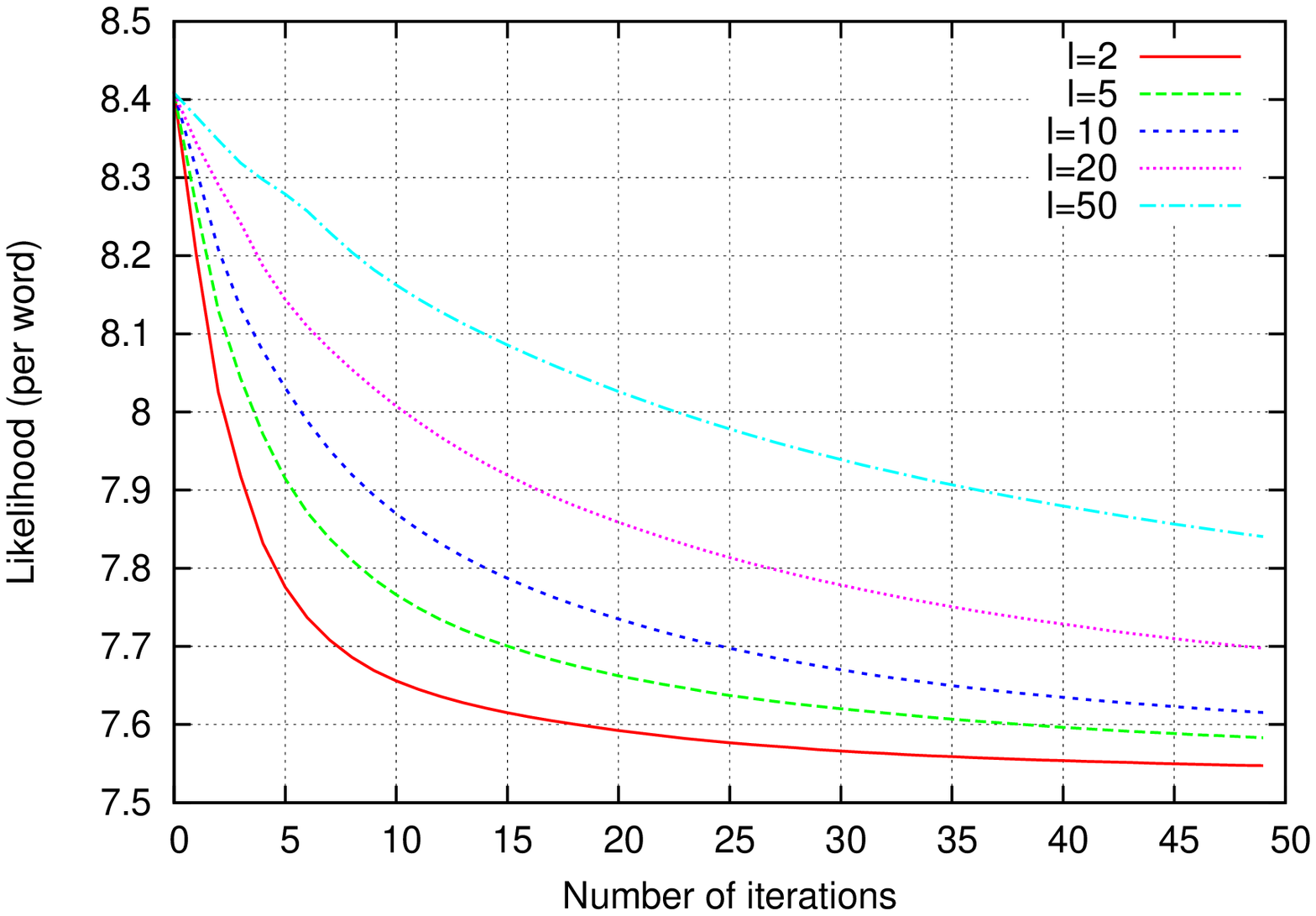}
  \caption{Convergence of the likelihood for the sparse sampler for
    LDA with various sparsity parameters}\label{lik3}
\end{figure}

Figures \ref{lik1}-\ref{lik3} show the convergence of the likelihood
and the AUC for BayesNet, for some combinations of LDA and linked LDA
models run with various Gibbs samplers. The plots range over 50
iterations, and we have stopped inference after every 2 iterations and
calculated the AUC of a BayesNet classification over the $\vartheta$
features, and the likelihood (as described in Subsections \ref{ss:lda}
and \ref{ss:llda}). The experiments are run on the small corpus, the
number of topics was $k=30$ for all models, and the parameters were as
described in Subsection \ref{ss:tlda}.

The high pairwise correlation of the otherwise quite different
accuracy measures, likelihood and AUC values in Figure \ref{lik1} is
very interesting. This is certainly due to the fact that after the
topic assignment stabilizes, there is only negligible variance in the
$\vartheta$ features. This behavior indicates that the widely accepted
method of stopping LDA iterations right after the likelihood has
stabilized can be used even if the inferred variables ($\vartheta$ in
our case) are later input to other classification methods.

The usual choice for the number of Gibbs sampling iterations for LDA
is 500-1000.  Thus it is worth emphasizing that in our experiments
after only 20-30 iterations both likelihood and accuracy stabilizes.
This is in accordance with the similar experiments of
\cite{griffiths2004fst,walker2008mbd}, which found that for plain LDA,
likelihood stabilizes after 50-100 iterations, over various corpora.
As a consequence, we chose 50 as the number of iterations for the next
experiments.

Figure \ref{lik2} demonstrates that the linked LDA model with plain,
aggregated and limit samplers over-perform plain LDA by about 1\% in
likelihood. This gap increases to about 4\% after applying classifiers
to the inferred $\vartheta$ topic distributions, see Table
\ref{tabsmall}. The aggregated and limit Gibbs boostings result in
negligible deterioration in the likelihood, though they give 5 times
speedup.

The observation that much fewer iterations are enough for Gibbs
sampling, combined with our Gibbs boosting methods bids fair that LDA
may become a computationally highly efficient latent topic model in
the future.

\subsection{Comparison with the baseline}\label{ss:baseline}
We run supervised web document classification as de\-scri\-bed in
Subsection \ref{ss:class}. The results can be seen in Tables
\ref{tabbig}-\ref{tabgr}, the evaluation metric is AUC, averaged over
the 11 big categories. 
%For linked LDA with limit Gibbs sampler one has
%to maintain a matrix of size $k \times |S_d|$ for every word in a
%document $d$, which required unfeasibly much space for the big
%corpus. This holds also for sparse Gibbs sampler, as it is built upon
%our limit sampler, thus, to avoid this obstacle, we measured the
%performance of the aggregated sparse sampler, which is the lazy
%version of aggregated Gibbs sampler.  
See Table \ref{tabbig} for AUC
values. The big corpus was too large for the baseline methods to
terminate, so comparison with them in the small corpus can be seen in
Table \ref{tabsmall}.

\begin{table}[ht]
\begin{center}
  \begin{tabular}{|l|c|c|c|}
    \hline
    model / sampler&BayesNet&SVM&C4.5\\
    \hline
    LDA          &0.821&0.710&0.762\\%simple
    LDA / aggr.  &0.820&0.684&0.756\\%cs
    LDA / limit  &0.810&0.695&0.739\\%csr
    LDA / sparse ($\ell=10$) &0.788&0.669&0.719\\
    \hline
    linked LDA   &\textbf{0.854}&\textbf{0.723}&\textbf{0.765}\\
    linked LDA / aggr.   &0.848&0.711&0.754\\
    linked LDA / a.~sparse ($\ell=10$) &0.837&0.701&0.733\\
    \hline
\end{tabular}
\caption{\label{tabbig} Big corpus. Classification accuracy measured
  in AUC for LDA and linked LDA under various Gibbs sampling
  heuristics. Sparse' at linked LDA refers to the lazy version of the
  aggregated Gibbs sampler.}
\end{center}
\end{table}

\begin{table}[ht]
\begin{center}
  \begin{tabular}{|l|c|c|c|}
    \hline
    model / sampler&BayesNet&SVM&C4.5\\
    \hline
    LDA          &0.817&0.705&0.767\\%simple
    LDA / aggr.  &0.813&0.691&0.750\\%cs
    LDA / limit  &0.808&0.662&0.720\\%csr
    LDA / sparse ($\ell=2$)&0.805&0.654&0.719\\
    LDA / sparse ($\ell=5$)&0.799&0.671&0.713\\
    LDA / sparse ($\ell=10$)&0.791&0.667&0.711\\
    LDA / sparse ($\ell=20$)&0.764&0.649&0.689\\
    LDA / sparse ($\ell=50$)&0.735&0.624&0.670\\
    \hline
    linked LDA         &\textbf{0.850}&0.709&\textbf{0.777}\\%linked
    linked LDA / aggr  &0.849&0.696&0.771\\%cl
    linked LDA / limit &0.845&0.688&0.761\\%clr
    linked LDA / sparse ($\ell=2$)&0.840&0.683&0.758\\
    linked LDA / sparse ($\ell=5$)&0.836&0.679&0.753\\
    linked LDA / sparse ($\ell=10$)&0.827&0.673&0.751\\
    linked LDA / sparse ($\ell=20$)&0.799&0.656&0.726\\
    linked LDA / sparse ($\ell=50$)&0.768&0.630&0.705\\
    \hline
    link-PLSA-LDA/var.~inf. \cite{nallapati:jlt}&0.827&0.687&0.754\\
    \hline
    tf.idf &0.569&\textbf{0.720}&0.565\\
    \hline
\end{tabular}
\caption{\label{tabsmall} Small corpus. Classification accuracy measured in AUC for LDA
  and linked LDA under various Gibbs sampling heuristics as well as
  the baseline methods.}
\end{center}
\end{table}

The tables clearly indicate that applying aggregated, limit and sparse
Gibbs sampling with sparsity $\ell$ at most 10 has only a minor
negative effect of about 2\% on the classification accuracy, albeit
they give significant speedup by Table \ref{tspeedup1}.  Linked LDA
slightly outperforms the LDA-based categorization for all classifiers,
by about 4\%.  This gap is biggest for BayesNet.

\begin{table}[ht]
\begin{center}
  \begin{tabular}{|l|c|c|c|}
    \hline
    model + graph&BNet&C4.5\\
    \hline
    LDA + cocit&0.830&0.762\\
    LDA + rev-cocit&0.831&0.770\\
    \hline
    linked LDA + $\chi$ &\textbf{0.863}&\textbf{0.785}\\
    linked LDA + rev-$\chi$ &0.857&0.780\\
    linked LDA + cocit &0.838&0.765\\
    linked LDA + rev-cocit &0.839&0.752\\
    \hline
    link-PLSA-LDA + own-$\chi$ &0.839&0.755\\
    link-PLSA-LDA + own-rev-$\chi$ &0.837&0.755\\
    link-PLSA-LDA + cocit &0.835&0.754\\
    link-PLSA-LDA + rev-cocit &0.840&0.763\\
    \hline
    tf.idf + cocit &0.597&0.589\\
    tf.idf + rev-cocit &0.595&0.593\\
    tf.idf + linked LDA-$\chi$ &0.611&0.601\\
    tf.idf + linked LDA-rev-$\chi$ &0.609&0.598\\
    tf.idf + link-PLSA-LDA-$\chi$ &0.604&0.597\\
    tf.idf + link-PLSA-LDA-rev-$\chi$ &0.606&0.596\\
    \hline
\end{tabular}
\caption{\label{tabgr} Small corpus, classification with graph
  stacking.  The base learner may be one of tf.idf, LDA, linked LDA
  (both without Gibbs sampling heuristics) and link-PLSA-LDA.  Edge
  weights may arise by cocitation or the inferred $\chi$ of the linked
  LDA and link-PLSA-LDA models.  These weights may be obtained over
  the reversed graph, which is indicated by \emph{rev}.}
\end{center}
\end{table}

Table \ref{tabgr} indicates that the $\chi$ link weights delivered by
the linked LDA model captures influence very well, as it improves 2\%
over tf.idf, 4\% over LDA and 3\% over linked LDA with the cocitation
graph, and 3\% over link-PLSA-LDA with its own $\chi$ weights. This
clearly indicates that the $\chi$ link weights provided by the linked
LDA model are good approximation of the topical similarity along
links.  Reversing the graph influences behaves in a quite
unpredictable way, though rev-cocit is somewhat better than cocit,
furthermore, reversion worsens the AUC measure if the weights come
from linked or link-PLSA-LDA $\chi$ values.

Every run (LDA model build and classification with and without graph
stacking) is repeated 10 times to get variance of the AUC measure.
Somewhat interestingly, this was at most $0.015$ throughout, so we
decided not to quote them individually.

\section{Conclusion and future work}\label{s:fut}
In this paper we introduced the linked LDA model which integrates the
flow of influence along links into LDA in such a way that each
document can be citing and cited at the same time. By our strategies
to boost Gibbs sampling we were able to apply our model to supervised
web document classification as a feature generation and dimensionality
reduction method.  In our experiments linked LDA outperformed LDA and
other link based LDA models by about 4\% in AUC.  One of our Gibbs
sampler heuristics produced 10-fold speedup with negligible
deterioration in convergence, likelihood and classification accuracy.
Over our data set of Web hosts, these boostings outperform the fast
Gibbs sampler of \cite{porteous2008fcg} in speed to a great extent.
We also note that our samplers use ideas orthogonal to fast Gibbs
sampling \cite{porteous2008fcg} and the paralleled sampling of
\cite{newman2007dil}, and so these methods can be used in combination.
It would be interesting to explore other domains than LDA where our
Gibbs sampling strategies can be applied.  Limit Gibbs sampling makes
it possible to have arbitrary non-negative real numbers as word counts
in a document, instead of the usual tf counts. To this end, we plan to
measure whether accuracy of LDA is improved if the tf counts are
replaced with the pivoted tf.idf counts of \cite{singhal1996dln}.  As
a further research we will investigate possible application of the
linked LDA model to other domains, like web spam filtering.

\subsection*{Acknowledgment} To Zoltán Gy\"ongyi for fruitful
discussions and for providing us with the DMOZ labels over the
UK2007-WEBSPAM corpus, and to Ramesh Nallapati for providing us their
link-PLSA-LDA and pairwise-link-LDA code.

\bibliographystyle{abbrv}
\bibliography{spam,www,lda}
%\bibliography{latex-common/bib/spam,latex-common/bib/www,latex-common/bib/lda}
\end{document}